\documentclass[usenatbib]{mnras}
\usepackage{amsmath}
\usepackage{natbib}
\usepackage{epsfig}
\usepackage{graphicx}
\usepackage{amssymb}
\usepackage{url}
\usepackage{xcolor}

\def\gtsima{$\; \buildrel > \over \sim \;$}
\def\ltsima{$\; \buildrel < \over \sim \;$}
\def\gtrsim{\lower.5ex\hbox{\gtsima}}
\def\lesssim{\lower.5ex\hbox{\ltsima}}

\newcommand{\msun}{$\mathrm{M}_{\odot}$}

\usepackage{amsmath}	
\usepackage{amssymb}	
\usepackage{mathrsfs}

\begin{document}

\title[BBHs in YSCs: the impact of metallicity]{Binary black holes in young star clusters: the impact of metallicity}
\author[]{Ugo N. Di Carlo$^{1,2,3}$, Michela Mapelli$^{4,2,3}$, Nicola Giacobbo$^{4,2,3}$,  Mario Spera$^{2,4,6,7}$, \newauthor Yann Bouffanais$^{4,2}$, Sara Rastello$^{4,2}$, Filippo Santoliquido$^{4,2}$, Mario Pasquato$^{2,3}$, \newauthor Alessandro Ballone$^{4,2}$, Alessandro A. Trani$^{8,9}$, Stefano Torniamenti$^{4,2}$, Francesco Haardt$^{1}$ 
\vspace{0.3cm}
\\
$^{1}$Dipartimento di Scienza e Alta Tecnologia, University of Insubria, Via Valleggio 11, I--22100, Como, Italy
\\
$^{2}$INFN, Sezione di Padova, Via Marzolo 8, I--35131, Padova, Italy
\\
$^{3}$INAF-Osservatorio Astronomico di Padova, Vicolo dell'Osservatorio 5, I--35122, Padova, Italy
\\
$^{4}$Dipartimento di Fisica e Astronomia `G. Galilei', University of Padova, Vicolo dell'Osservatorio 3, I--35122, Padova, Italy
\\
$^{5}$Scuola Internazionale Superiore di Studi Avanzati (SISSA), Via Bonomea 265, I-34136 Trieste, Italy
\\
$^{6}$Center for Interdisciplinary Exploration and Research in Astrophysics (CIERA), Evanston, IL 60208, USA
\\
$^{7}$Department of Physics \& Astronomy, Northwestern University, Evanston, IL 60208, USA \\
$^{8}$Department of Astronomy, Graduate School of Science, The University of Tokyo, 7-3-1 Hongo, Bunkyo-ku, Tokyo, 113-0033, Japan \\
$^{9}$Department of Earth Science and Astronomy, College of Arts and Sciences, The University of Tokyo, 3-8-1 Komaba, Meguro-ku,\\ Tokyo 153-8902, Japan \\
}
\maketitle \vspace {7cm}
\bibliographystyle{mnras}

\begin{abstract}
Young star clusters are the most common birth-place of massive stars and are dynamically active environments. Here, we study the formation of black holes (BHs) and binary black holes (BBHs) in young star clusters, by means of 6000 N-body simulations coupled with binary population synthesis. We  probe three different stellar metallicities ($Z=0.02, 0.002$ and 0.0002) and two initial density regimes (density at the half-mass radius $\rho_{\rm h}\ge{}3.4\times10^4$ and $\ge{1.5\times10^2}$  M$_\odot$ pc$^{-3}$ in dense and loose star clusters, respectively). Metal-poor clusters tend to form more massive BHs than metal-rich ones. We find $\sim{}6$, $\sim{}2$, and $<1$~\%  of BHs with mass $m_{\rm BH}>60$ M$_\odot$ at $Z=0.0002,$ 0.002 and 0.02, respectively. In metal-poor clusters, we form intermediate-mass BHs with mass up to $\sim{}320$ M$_\odot$. BBH mergers born via dynamical exchanges (exchanged BBHs) can be more massive than BBH mergers formed from binary evolution: the former (latter) reach total mass up to $\sim{}140$ M$_\odot$ ($\sim{}80$ M$_\odot$). The most massive BBH merger in our simulations has primary mass $\sim{}88$ M$_\odot$, inside the pair-instability mass gap, and a mass ratio of $\sim{}0.5$. Only BBHs born in young star clusters from metal-poor progenitors can match the masses of GW170729, the most massive event in O1 and O2, and those of GW190412, the first unequal-mass merger. We estimate a local BBH merger rate density $\sim{}110$ and $\sim{}55$ Gpc$^{-3}$ yr$^{-1}$, if we assume that all stars form in loose and dense star clusters, respectively.
\end{abstract}

\begin{keywords}
black hole physics -- gravitational waves -- methods: numerical -- galaxies: star clusters: general -- stars: kinematics and dynamics -- binaries: general 
\end{keywords}

\maketitle

%

\section{Introduction}
About four years ago, the LIGO-Virgo collaboration (LVC, \citealt{LIGOdetector,Virgodetector}) reported the very first direct detection of gravitational waves, GW150914, interpreted as the merger of two massive stellar black holes (BHs, \citealt{abbottGW150914,abbottastrophysics}). After GW150914, nine additional binary black holes (BBHs) and one binary neutron star (BNS) were observed by the LVC during the first and second observing run (hereafter O1 and O2, \citealt{abbottO1,abbottO2,abbottO2popandrate}). The third observing run of LIGO and Virgo 
has recently been completed and has already led to one additional BNS (GW190425, \citealt{abbottGW190425}), the first unequal-mass BBH merger (GW190412, \citealt{abbottGW190412})  and tens of public alerts\footnote{https://gracedb.ligo.org/}.

Understanding the formation channels of BBHs is one of the most urgent astrophysical questions raised by LVC observations. Several authors suggest that about a hundred of detections are sufficient to say something on the formation channels of BBHs, thanks to their distinctive signatures (e.g. \citealt{fishbach2017, gerosa2017, stevenson2017, gerosa2018,bouffanais2019}).

Isolated binary evolution, either via common envelope (e.g. \citealt{tutukov1973, bethe1998,portegieszwart1998,belczynski2002, voss2003, podsiadlowski2004, belczynski2008,dominik2012, dominik2013,  mennekens2014, belczynski2016,mapelli2017,mapelli2018,giacobbo2018, giacobbo2018b, kruckow2018,spera2018,mapelli2019,neijssel2019,tang2019}) or via chemically homogeneous scenarios \citep{demink2016,mandel2016,marchant2016}, predicts the formation of BBHs with primary mass up to $\sim{}40-65$ M$_\odot$ (see, e.g., \citealt{mapelli2020} and references therein), with a strong preference for equal-mass systems, mostly aligned spins and zero eccentricity in the LVC band.

In contrast, dynamical formation in star clusters might lead to even larger primary masses (e.g. \citealt{mckernan2012,mapelli2016,antonini2016,gerosa2017,stone2017,mckernan2018,dicarlo2019,dicarlo2020,rodriguez2019,yang2019,arcasedda2019,arcasedda2020}), mass ratios ranging from $q\sim{}0.1$ to $q\sim{}1$ (e.g. \citealt{dicarlo2019}), isotropic spin distribution, and, in some rare but not negligible cases, non-zero eccentricity in the LVC band (e.g. \citealt{samsing2018,samsing2018b,samsing2018c,rodriguez2018,zevin2019}).

The zoology of star clusters found in the Universe is rich and includes systems that are extremely different from each other (both in terms of mass and lifetime), but share a similar dynamical evolution: almost all star clusters are collisional systems, i.e. stellar systems in which the two-body relaxation timescale is shorter than their lifetime (e.g. \citealt{spitzer1987}). Hence, close encounters between single and binary (or multiple) stars drive the evolution of star clusters and have a crucial impact on the formation of binary compact objects.

The dynamical evolution of BBHs in nuclear star clusters  (e.g. \citealt{oleary2009,millerlauburg2009,mckernan2012,mckernan2018,vanlandingham2016,stone2017,hoang2018,arcagualandris2018,antonini2018}) and  globular clusters (e.g. \citealt{sigurdsson1993,sigurdsson1995,portegieszwart2000,oleary2006,sadowski2008,downing2010,downing2011,tanikawa2013,rodriguez2015,rodriguez2016,rodriguez2018,antonini2016,antonini2018,hurley2016,oleary2016,askar2017,askar2018,zevin2017,choksi2019}) has been extensively investigated. These are the most massive, long-lived and predominantly old stellar systems; hence their relatively high escape velocity allows a fraction of the merger remnants to stay in the cluster, leading to a population of hierarchical mergers \citep{miller2002,arcasedda2019,gerosa2019,rodriguez2019}.

Young star clusters (YSCs) and open clusters are generally smaller and  shorter-lived than globular clusters \citep{portegieszwart2010}. Nonetheless, they are site of strong dynamical interactions and they are the nursery of massive stars in the Universe: the vast majority of massive stars, which are the progenitors of compact objects, form in YSCs (e.g. \citealt{lada2003,portegieszwart2010}). Hence, the majority of BHs have likely spent the first part of their life in star clusters, undergoing dynamical encounters. Several studies demonstrate that dynamics has a major role in the formation of BH binaries in YSCs \citep{portegieszwart2002, banerjee2010,mapelli2013,mapelli2014,ziosi2014, goswami2014, mapelli2016, banerjee2017, banerjee2018, fujii2017, rastello2018,dicarlo2019,kumamoto2019,kumamoto2020}.

In particular, \cite{dicarlo2019} showed that about half of BBHs born in YSCs form via dynamical exchanges at metallicity $Z=0.002$. BBHs formed in YSCs are significantly more massive than BBHs formed from isolated binary evolution and tend to have smaller mass ratios. About $\sim{}2$~\% of all BBH mergers originating from YSCs have primary mass $\gtrsim{}60$ M$_\odot$, falling inside the pair-instability mass gap (e.g. \citealt{woosley2017,spera2017,stevenson2019,marchant2019,farmer2019,mapelli2020,dicarlo2020,renzo2020}). The sample presented in \cite{dicarlo2019} is the largest simulation set of YSCs used to study BBHs, but is limited to one metallicity $Z=0.002$. Since metallicity has a crucial impact on the mass of BHs \citep{mapelli2009,mapelli2010,zampieri2009,belczynski2010,spera2015}, it is essential to study the evolution of BBHs in star clusters with different metallicity. In this paper, we present the result of a new set of simulations where we consider three different metallicities ($Z=0.02$, 0.002 and 0.0002) and two initial density regimes (density at the half-mass radius $\rho_{\rm h}\ge{}3.4\times10^4$ and $\ge{1.5\times10^2}$  M$_\odot$ pc$^{-3}$ in dense and loose star clusters, respectively).

\section{Methods}
The simulations discussed in this paper were done using the same code and methodology as described in \cite{dicarlo2019}. In particular, we use the direct summation N-Body code \textsc{nbody6++gpu} \citep{wang2015} coupled with the population synthesis code \textsc{mobse} \citep{mapelli2017,giacobbo2018,giacobbo2018b}.

\subsection{Direct N-Body}

\textsc{nbody6++gpu} is the GPU parallel version of \textsc{nbody6} \citep{aarseth2003}. It implements a 4th-order Hermite integrator, individual block time–steps \citep{makino1992} and Kustaanheimo-Stiefel (KS) regularization of close encounters and few-body subsystems \citep{stiefel1965,kschain}.

A neighbour scheme \citep{nitadori2012} is used to compute the force contributions at short time intervals (\textit{irregular} force/timesteps), while at longer time intervals (\textit{regular} force/timesteps) all the members in the system contribute to the force evaluation. The irregular forces are evaluated using CPUs, while the regular forces are computed on GPUs using the CUDA architecture.
This version of \textsc{nbody6++gpu} does not include post-Newtonian terms. 

\subsection{Population synthesis}

\textsc{mobse} \citep{mapelli2017,giacobbo2018,giacobbo2018b,giacobbo2018c,mapelli2018} is a customized and upgraded version of {\sc bse} \citep{hurley2000,hurley2002} which includes up-to-date prescriptions for massive star winds, for core-collapse supernova (SN) explosions and for pair instability and pulsational-pair instability SNe. It has been integrated with \textsc{nbody6++gpu} by taking advantage of the pre-existing interface between the N-body code and {\sc bse}.

Stellar winds are implemented assuming that the mass loss of massive hot stars (O and B-type stars, Wolf-Rayet stars, luminous blue variable stars) depends on metallicity as  $\dot{M}\propto Z^\beta$, where $\beta$ is defined as in \cite{giacobbo2018}
\begin{equation}
\beta = \begin{cases}
0.85 & \mathrm{if} \quad \Gamma_e < 2/3\\
2.45-2.4\Gamma_e & \mathrm{if} \quad 2/3 \leq \Gamma_e < 1 \\
0.05 & \mathrm{if}\quad \Gamma_e\geq 1.
\end{cases}
\end{equation}
Here $\Gamma_e$ is the Eddington factor (see e.g. \citealt{graefener2008,chen2015}).



The outcome of core-collapse SNe is highly uncertain and none of the prescriptions available in the literature is completely satisfactory (e.g. \citealt{burrows2018,mapelli2020}). Hence, our prescriptions should be regarded as reasonable ``toy models''. In this paper, we adopt the rapid core-collapse supernova model described in \cite{fryer2012}. In this formalism, the mass of the compact object is $m_{\rm rem} = m_{\rm proto} + m_{\rm fb}$, where $m_{\rm  proto} = 1$ M$_\odot$ is the mass of the proto-compact object and $m_{\rm fb}$ is the mass accreted by fallback. Note that this is different from \cite{dicarlo2019}, where we adopted the delayed model from \cite{fryer2012}.

When the helium core of a star becomes $64\le{}m_{\rm He}/{\rm M}_{\odot}\le{}135$, the star is completely destroyed by pair instability. If the helium core reaches a size $32\le{}m_{\rm He}/{\rm M}_{\odot}<64$, pulsational pair instability is expected to take place \citep{woosley2017} and the final mass of the compact object is estimated as $m_{\rm rem} = \alpha_{\rm P}\,{} m_{\rm no\,{} PPI}$, where $m_{\rm no\,{} PPI}$ is the mass of the compact object we would have obtained if we had not included pulsational pair instability in our analysis and $\alpha{}_{\rm P}$ is a fitting parameter \citep{spera2017,mapelli2020}. Finally, electron-capture supernovae are implemented as described in \cite{giacobbo2018c}.

Natal kicks are randomly drawn from a Maxwellian velocity distribution. A one-dimensional root mean square velocity $\sigma{}=15$ km s$^{-1}$ is adopted for core-collapse SNe and for electron-capture SNe \citep{giacobbo2018c}. Kick velocities of BHs are reduced by the amount of fallback as $V_{\mathrm{KICK}}=(1-f_{\mathrm{fb}})\,{}V$, where $f_{\mathrm{fb}}$ is the fallback parameter described in \cite{fryer2012} and $V$ is the velocity drawn from the Maxwellian distribution\footnote{This kick model was chosen because it leads to a BNS merger rate in agreement with the range inferred from the LVC \citep{baibhav2019}, but is in tension with the proper motions of young Galactic pulsars \citep{hobbs2005}. In a recent work \citep{giacobbo2020}, we have revised our kick prescriptions  and we have shown that the value of $\sigma{}$ adopted in this work has negligible effect on the properties and on the merger rate of BBHs (because $V_{\rm KICK}$ is dominated by fallback).}. 

Binary evolution processes (tides, mass transfer, common envelope and gravitational-wave orbital decay) are implemented as in \cite{hurley2002}. In this work, we assume $\alpha{}=5$ (it was $\alpha{}=3$ in \citealt{dicarlo2019}), while $\lambda{}$ is derived by \textsc{mobse} as described in \cite{claeys2014}.


Consistently with \cite{dicarlo2019}, when two stars merge, the amount of mass loss is decided 
by \textsc{mobse}, which adopts the same prescriptions as \textsc{bse}, but  if a star merges with a BH or a neutron star, \textsc{mobse} assumes that the entire mass of the star is immediately lost by the system and the compact object does not accrete it. 
This assumption by \textsc{mobse} is very conservative, because it is unlikely that the compact object can accrete a substantial fraction of the stellar mass, but it is hard to quantify the actual mass accretion.

\begin{table*}
\begin{center}
\caption{\label{tab:table1} Initial conditions.} \leavevmode
\begin{tabular}[!h]{ccccc}
\hline
Set & $Z$ & $N_{\rm SC}$ & $M_{\rm SC}$ [M$_\odot{}$] & $r_{\rm h}$ [pc]
\\
\hline
YSC & 0.02, 0.002, 0.0002 & 6000 &  $10^3-3\times 10^4$ &  1.5, $0.1\,{} \left( M_{\mathrm{SC}}/{\rm M}_{\odot}\right)^{0.13}$ \\
A & 0.02, 0.002, 0.0002 & 3000 &  $10^3-3\times 10^4$ &  $0.1\,{} \left( M_{\mathrm{SC}}/{\rm M}_{\odot}\right)^{0.13}$\\
B & 0.02, 0.002, 0.0002 & 3000 &  $10^3-3\times 10^4$ &  1.5\\
IB & 0.02, 0.002, 0.0002        & $3\times 10^7$ & -- & -- \\
\hline
A02 & 0.02 & 1000 &  $10^3-3\times 10^4$ &  $0.1\,{} \left( M_{\mathrm{SC}}/{\rm M}_{\odot}\right)^{0.13}$\\
A002 & 0.002 & 1000 &  $10^3-3\times 10^4$ &  $0.1\,{} \left( M_{\mathrm{SC}}/{\rm M}_{\odot}\right)^{0.13}$\\
A0002 & 0.0002 & 1000 &  $10^3-3\times 10^4$ &  $0.1\,{} \left( M_{\mathrm{SC}}/{\rm M}_{\odot}\right)^{0.13}$\\

B02 & 0.02 & 1000 &  $10^3-3\times 10^4$ &  1.5\\
B002 & 0.002 & 1000 &  $10^3-3\times 10^4$ &  1.5\\
B0002 & 0.0002 & 1000 &  $10^3-3\times 10^4$ &  1.5\\

IB02 & 0.02 & $10^7$ & -- & -- \\
IB002 & 0.002 & $10^7$ & -- & -- \\
IB0002 & 0.0002 & $10^7$ & -- & -- \\
\hline
\end{tabular}
\end{center}
\begin{flushleft}
\footnotesize{Column~1: Name of the simulation set; YSC stands for all dynamical simulations (set A and set B) considered together, while IB stands for isolated binaries. Column~2 ($Z$): stellar metallicity; column~3 ($N_{\rm SC}$): Number of runs; column~4: YSC mass ($M_{\rm SC}$); column~5: initial half-mass radius ($r_{\mathrm{h}}$).} 
\end{flushleft}
\end{table*}

\begin{figure}
  \center{
    \epsfig{figure=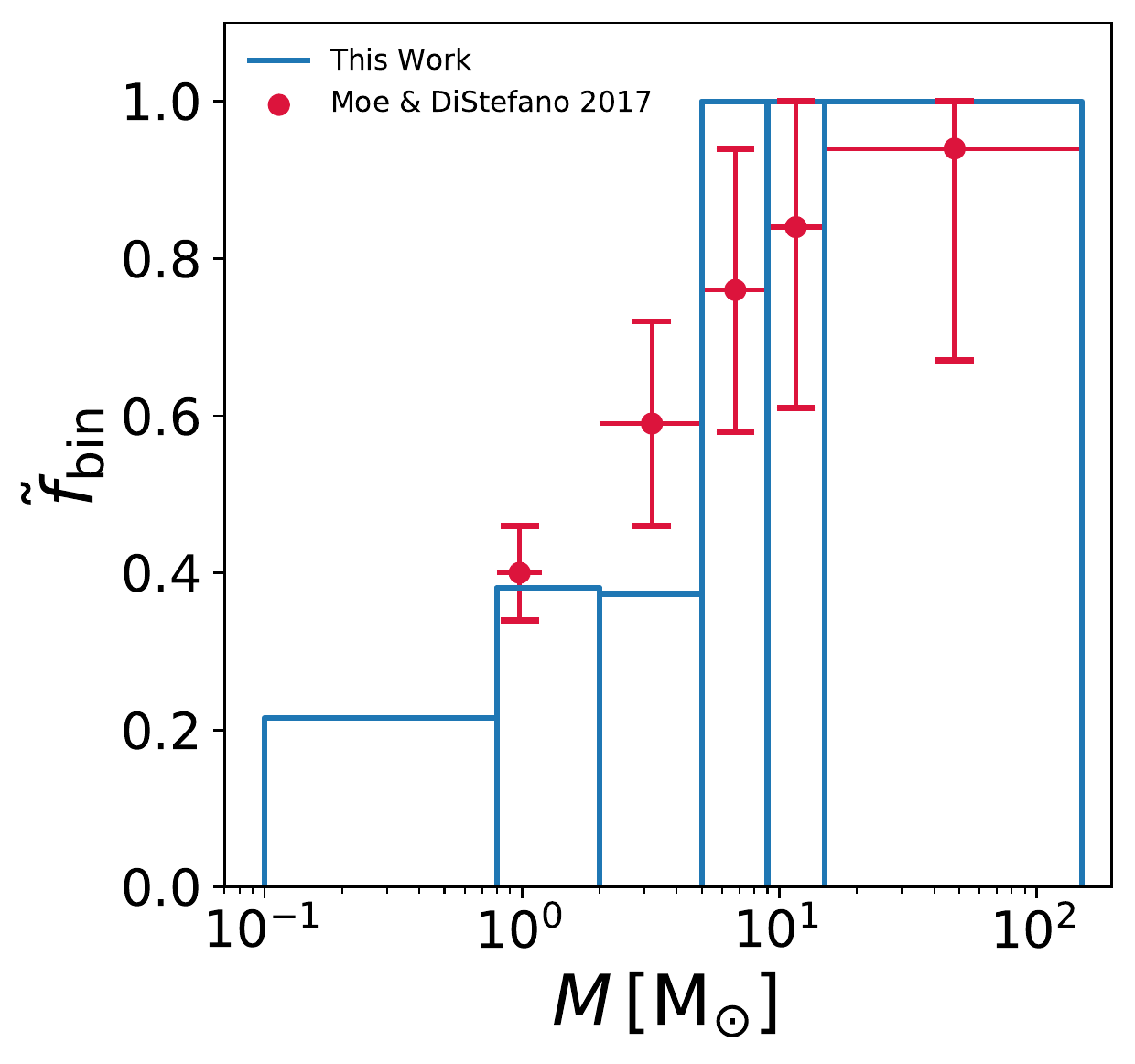,width=8.0cm}
    \caption{\label{fig:binfrac}
Initial binary fraction $\widetilde{f}_{\mathrm{bin}}$ as a function of stellar mass. $\widetilde{f}_{\mathrm{bin}}$ is defined as $N_{\mathrm{bin}}/(N_{\mathrm{bin}}+N_{\mathrm{sin}})$, where $N_{\mathrm{bin}}$ is the total number of binaries and $N_{\mathrm{sin}}$ is the total number of single stars in the YSC at the beginning of the simulation. The blue line represents the binary fraction for one of our simulated star clusters, while the red circles come from the observational results \citep{moe2017} and represent the fraction of stars with at least one companion.
}}
\end{figure}

\subsection{Initial conditions}
We have simulated 6000 YSCs considering three different metallicities ($Z=0.02, 0.002,$ and 0.0002) and two definitions for the initial half-mass radius $r_{\rm h}$ (Table~\ref{tab:table1}). Simulations of set~A (3000 simulations, 1000 per each considered metallicity) were performed choosing $r_{\rm h}$ according to the Marks \& Kroupa relation \citep{marks2012}, which relates the total mass $M_{\mathrm{SC}}$ of a star cluster at birth with its initial half mass radius $r_{\rm h}$:
\begin{equation}
r_{\rm h}=0.10^{+0.07}_{-0.04}\,{}{\rm pc}\,{} \left( \frac{M_{\mathrm{SC}}}{M_{\odot}}\right)^{0.13\pm 0.04}.
\end{equation}

Simulations of set~B (3000 simulations, 1000 per each considered metallicity) assume $r_{\rm h}=1.5\,\mathrm{pc}$. The initial densities of the YSCs at the half-mass radius are $\rho_{\rm h}=500\left( M_{\mathrm{SC}}/M_{\odot}\right)^{0.61}\,\rm{M}_{\odot}\rm{pc}^{-3}$ and $4/27 \left( M_{\mathrm{SC}}/M_{\odot}\right)\,\rm{M}_{\odot}\rm{pc}^{-3}$ for set A and B, respectively. We also refer to set A/set B SCs as dense/loose ones.

As already discussed in \cite{dicarlo2019}, we model YSCs with fractal initial conditions, because this mimics the initial clumpiness and asymmetry of embedded star clusters \citep{cartwright2004,gutermuth2005,goodwin2004, ballone2020}. We adopt a fractal dimension $D=1.6$ and generate the initial conditions with  \textsc{McLuster}  \citep{kuepper2011}. In \cite{dicarlo2019}, we have shown that larger values of the fractal dimension ($D\le{}2.3$) do not significantly affect the statistics of BBHs.

The total mass $M_{\rm SC}$ of each star cluster (ranging from $1000$ \msun{} to $30000$ \msun{})  is drawn from a distribution $dN/dM_{\rm SC}\propto M_{\rm SC}^{-2}$, as the embedded star cluster mass function described in \cite{lada2003}. Thus, the mass distribution of our simulated star clusters mimics the mass distribution of star clusters in Milky Way-like galaxies. The star clusters are initialised so that the virial ratio $\alpha_{\mathrm{vir}} = T/|V| = 0.5$, where $T$ and $V$ are the  total kinetic and potential energy of the YSC, respectively.


The stars in the simulated star clusters follow a \cite{kroupa2001} initial mass function, with minimum mass 0.1 \msun{}  and maximum mass 150 \msun{}. We assume an initial binary fraction $f_{\mathrm{bin}}=0.4$, meaning that $40$\% of the stars are members of binary systems. The orbital periods, eccentricities and mass ratios of binaries with primary more massive than 5 \msun{} are drawn from \cite{sana2012} distributions, as already described in \cite{dicarlo2019}. Stars with a mass larger than 5 \msun{}, starting from the most massive, are paired with the star which better matches the mass ratio drawn from the distribution. Stars under 5 \msun{} are randomly paired until the required binary fraction is reached. This procedure results in a mass-dependent initial binary fraction which is larger for more massive binaries, consistent with the multiplicity properties of O/B-type stars (e.g. \citealt{sana2012,moe2017}), as shown in Figure \ref{fig:binfrac}.


The force integration includes a solar neighbourhood-like static external tidal field \citep{wang2016}.  Each star cluster is evolved until its dissolution or for a maximum time $t=100\,\mathrm{Myr}$. The most massive star clusters in our sample are not completely disrupted at $t=100$ Myr, but our static tidal field model tends to overestimate the lifetime of star clusters, because it does not account for massive perturbers (e.g. molecular clouds), which can accelerate star cluster disruption \citep{gieles2006}. Hence, our choice is quite conservative. When the $N-$body simulation stops, we extract all the BBHs and we evolve their semi-major axis and eccentricity using the timescale formula presented in \cite{peters1964}, which describes the evolution of the orbit due to GW emission. We classify as merging BBHs all BBHs that merge within a Hubble time ($t_{\rm H}=14$ Gyr) by gravitational wave decay.

For comparison, we have also run a set of isolated binary simulations with the stand-alone version of {\sc mobse}. In particular, we simulated $10^7$ isolated binaries (IBs) per each considered metallicity ($Z=0.02,\,{}0.002$ and 0.0002). Primary masses of the IBs are drawn from a Kroupa \citep{kroupa2001} mass function between 5 and 150 M$_\odot$. Orbital periods and eccentricities are randomly drawn from the same distribution as the dynamical simulations, but for one difference: the maximum orbital period is $\log{(P_{\rm max}/{\rm days})}=5.5$ and $\log{(P_{\rm max}/{\rm days})}=6.7$ in the isolated binaries and in the dynamical simulations, respectively. We checked that this difference has a negligible impact on our results.  A summary of the initial conditions of the performed simulations is reported in Table \ref{tab:table1}.

\section{Results}
\begin{figure*}
  \center{
    \epsfig{figure=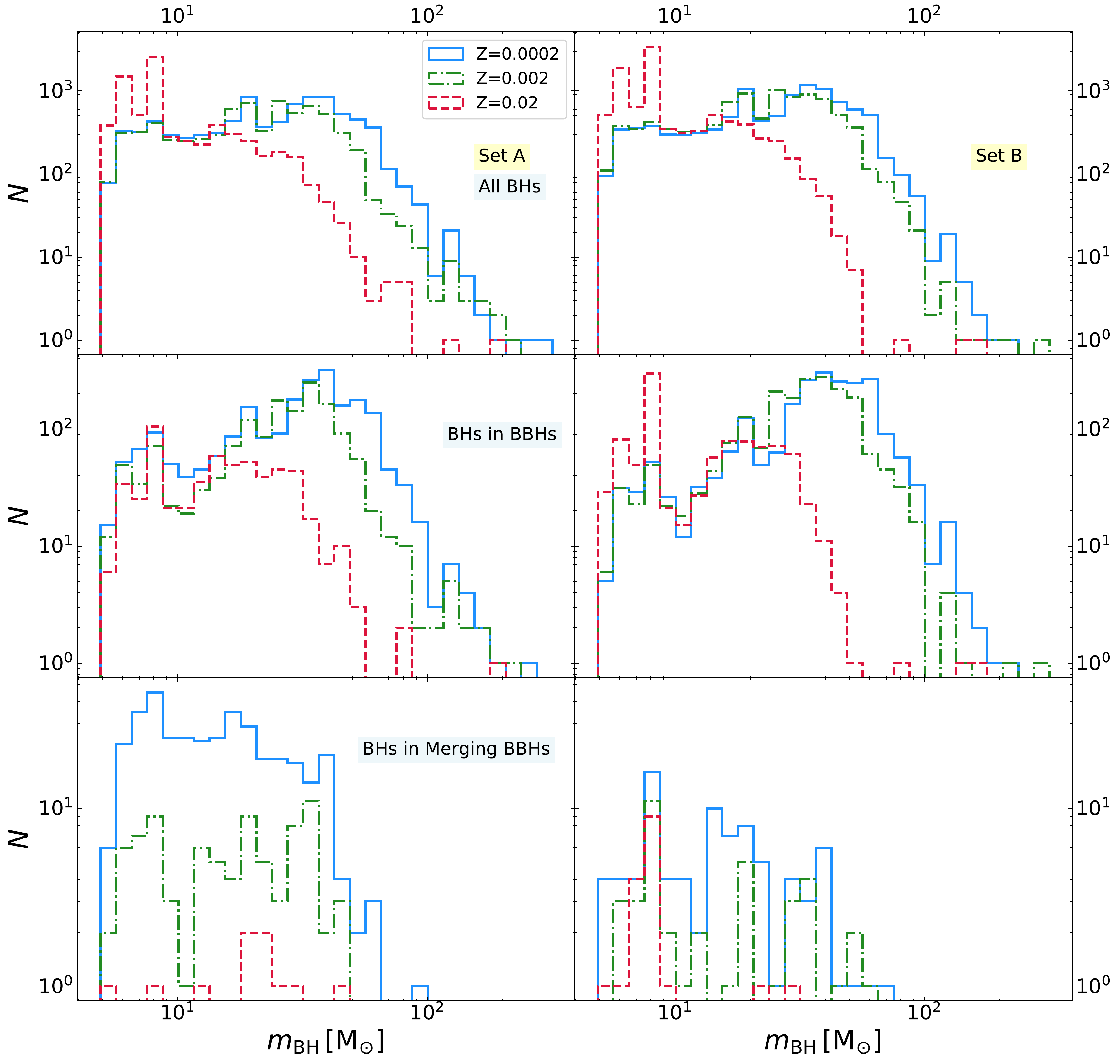,width=16.0cm}
    \caption{\label{fig:mhist}
    Distribution of  BH masses in the simulations. Left-hand panels: set~A; right-hand panels: set~B; top: all BHs; middle: BHs which are members of BBHs at the end of the simulations; bottom: BHs in merging BBHs. Blue solid line: $Z=0.0002$; green dot-dashed line: $Z=0.002$; red dashed line: $Z=0.02$.}}
\end{figure*}

\subsection{BH mass distribution}
Figure~\ref{fig:mhist} shows the mass distribution of all simulated BHs. The overall mass range of BHs, considering both single and binary BHs, spans from 5 M$_\odot$ (the minimum BH mass according to the rapid model by \citealt{fryer2012}) to 320 M$_\odot$. 

The maximum BH mass and the slope of the BH mass function depend on metallicity: BHs born from metal-rich stars ($Z=0.02$) tend to be less massive than BHs born from metal-poor stars ($Z=0.0002-0.002$). 

In the case of single stars and isolated binaries, {\sc mobse} predicts a maximum BH mass of $\sim{}65$ M$_\odot$ (see Figure~4 of \citealt{giacobbo2018}), while in our dynamical simulations we find BHs with mass up to $\sim{}320$ M$_\odot$. This difference is a result of multiple stellar mergers in YSCs, which build up a significantly more massive BH population in star clusters than in the field. This produces a non negligible population of BHs with mass in the pair-instability gap, between $\sim{}60$ and $\sim{}120$ M$_\odot$: $\sim{}5.0$~\%, $1.5$~\% and $0.2$~\% ($\sim{}5.7$~\%, $\sim{}2.2$~\% and $\sim{}0.01$~\%) of the simulated BHs have mass in the pair-instability gap in our set~A (set~B) at $Z=0.0002,$ 0.002 and 0.02, respectively.

Intermediate-mass BHs (IMBHs), defined as BHs with $m_{\rm BH}\ge{}100$~M$_\odot$, are   $\sim{}0.5$~\%, $\sim{}0.3$~\% and $\sim{}0.03$~\% ($\sim{}0.4$~\%, $\sim{}0.1$~\% and $\sim{}0.02$~\%) of all our BHs in set A (set B) at $Z=0.0002,$ 0.002 and 0.02, respectively.  They form through (multiple) stellar mergers, whose probability is enhanced by the short dynamical friction timescale in our clusters ($t_{\rm df}\lesssim{}1$ Myr for a star with zero-age main-sequence mass $m_{\rm ZAMS}\gtrsim{}20$ M$_\odot$): the most massive stars and binary stars sink to the core of the cluster before they become BHs; once in the core, they interact with each other triggering the mechanism known as runaway collision (e.g. \citealt{portegieszwart2002,portegieszwart2004,giersz2015,mapelli2016}).

The mass distribution of BHs in dense clusters (set A) and loose clusters (set B) are similar. The main difference is the percentage of BBHs that merge within a Hubble time (hereafter, merging BBHs), especially at low metallicity: these are $\sim{}17.1$~\%, $\sim{}5.7$~\% and $\sim{}1.7$~\% ($\sim{}3.9$~\%, $\sim{}2.0$~\% and $\sim{}1.8$~\%) in set A (set B) for a progenitor metallicity $Z=0.0002$, 0.002 and 0.02, respectively. Hence, star cluster density plays an important role in shrinking the orbit of BBHs. 
From these numbers, it is also apparent that  BBH mergers  are more common at low metallicity.




\begin{table}
\begin{center}
\caption{\label{tab:table2} Percentage of original and exchanged BBHs.} \leavevmode
\begin{tabular}[!h]{ccccc}
\hline
Set & $f_{\mathrm{orig, all}}$ & $f_{\mathrm{exch, all}}$ & $f_{\mathrm{orig, merger}}$ & $f_{\mathrm{exch, merger}}$
\\
\hline
YSC & 22 \% & 78 \% &  58 \% & 42 \% \\
A & 18 \% & 82 \% &  58 \% & 42 \% \\
B & 25 \% & 75 \% & 65 \% & 35 \%\\
\hline
A02 & 7 \% & 93 \% & 0 \% &  100 \% \\
A002 & 15 \% & 85 \% & 36 \% &  64 \%\\
A0002 & 24 \% & 76 \% & 65 \% &  35 \%\\

B02 & 28\% & 72 \% &  67 \% &  33 \% \\
B002 & 22 \% & 78 \% & 75 \% & 25 \%\\
B0002 & 25 \% & 75 \% & 60 \% & 40 \%\\

\hline
\end{tabular}
\end{center}
\begin{flushleft}
\footnotesize{Column~1: Name of the simulation set; column~2: $f_{\mathrm{orig, all}}$, percentage of original BBHs with respect to all BBHs at the end of the simulations; column~3 $f_{\mathrm{exch, all}}$, percentage of exchanged BBHs with respect to all BBHs at the end of the simulations; column~4: $f_{\mathrm{orig, merge}}$, percentage of merging original BBHs with respect to all merging BBHs; column~5: $f_{\mathrm{exch, merge}}$, percentage of merging exchanged BBHs with respect to all merging BBHs.}
\end{flushleft}
\end{table}

\subsection{Properties of merging BBHs}
Here, we focus on merging BBHs, i.e. BBHs that reach coalescence within a Hubble time. We call dynamical BBHs and isolated BBHs those merging BBHs that form in YSCs and in isolated binaries, respectively. We further divide dynamical BBHs into exchanged BBHs (i.e. dynamical BBHs that form from dynamical exchanges) and original BBHs (i.e. dynamical BBHs that form from binary stars which were already present in the initial conditions\footnote{In papers about star cluster dynamics, original BBHs are usually referred to as `primordial BBHs' or `BBHs born from primordial binaries', because the binary stars which were already present in the initial conditions are usually called `primordial binaries'. Here, we name them original BBHs to avoid confusion with primordial BHs that might form from gravitational instabilities in the early Universe (e.g. \citealt{carr1974,carr2016}).}).

Table~\ref{tab:table2} shows the percentage of original and exchanged BBHs for each set. About 78\% of all BBHs are exchanged, but only $\sim{}43$\% of the merging BBHs are exchanged. This indicates that a large fraction of exchanged BBHs are loose binaries and cannot harden fast enough to merge within a  Hubble time. 
The percentage of exchanged BBHs in set A is higher than that of set B: binaries in dense star clusters undergo more exchanges than in loose star clusters. 

The fraction of exchanged BBHs increases with metallicity in set~A, while it is almost constant with metallicity in set~B.  For example, the percentages of exchanged BBHs  and merging exchanged BBHs are $\sim{}76$ \%  and $\sim{}35$ \% in set A0002, and rise to   $\sim{}93$ \% and 100 \% in set A02. In contrast,  the percentages of exchanged BBHs  and merging exchanged BBHs are $\sim{}75$ \%  and $\sim{}40$ \% in set B0002, and remain very similar ($\sim{}72$ \% and 33 \%) in set B02. 

\begin{table*}
\begin{center}
\setlength{\tabcolsep}{15pt}
\caption{\label{tab:stat} Results of the KS-Test and U-Test to compare sets of merging BBHs.} \leavevmode
\begin{tabular}[!h]{lllll}
\hline
\textbf{Set 1} & \textbf{Set 2} & \textbf{Distribution} & \textbf{KS-Test} & \textbf{U-Test} \\
\hline
%
A -- Original & B -- Original & $m_{\rm tot}$ & 0.82 & 0.56 \\
A -- Exchanged & B -- Exchanged & $m_{\rm tot}$ & 0.36 & 0.55 \\
A -- All & B -- All & $m_{\rm tot}$ & 0.65 & 0.50 \vspace{0.2cm}\\

A -- Original & B -- Original & $m_{\rm chirp}$ & 0.57 & 0.35 \\
A -- Exchanged & B -- Exchanged & $m_{\rm chirp}$ & 0.56 & 0.59 \\
A -- All & B -- All & $m_{\rm chirp}$ & 0.33 & 0.38 \vspace{0.2cm}\\ 

A -- Original & B -- Original & $q$ & 0.05 & 0.14 \\
A -- Exchanged & B -- Exchanged & $q$ & 0.84 & 0.59 \\
A -- All & B -- All & $q$ & 0.50 & 0.54 \vspace{0.2cm}\\

A -- Original & B -- Original & $t_{\rm delay}$ & 0.43 & 0.35 \\
A -- Exchanged & B -- Exchanged  & $t_{\rm delay}$ & 0.99 & 0.94 \\
A -- All & B -- All & $t_{\rm delay}$ & 0.88 & 0.50 \\
\hline
\setlength{\tabcolsep}{6pt}
\end{tabular}
\end{center}
\begin{flushleft}
  \footnotesize{In this Table, we apply the KS- and U- tests to compare different samples of BBHs. Columns~1 and 2: the two BBH samples to which we apply the KS- and U- test. Each sample comes from one of the simulation sets (see Table~\ref{tab:table1}).
    Column~3: distribution to which we apply the KS- and U- tests. We consider total BBH masses ($m_{\rm tot}$), chirp masses ($m_{\rm chirp}$), mass ratios ($q$) and delay times ($t_{\rm delay}$).
    Columns~4 and 5: probability that the two samples are drawn from the same distribution according to the Kolmogorov-Smirnov (KS) Test and to the U-Test, respectively.}
\end{flushleft}
\end{table*}

Table~\ref{tab:stat} shows the results of the Kolmogorov-Smirnov (hereafter, KS) test \citep{birnbaum1951,wang2003} and of the U-test \citep{bauer1972,wolfe1999}. We find that the masses of merging BBHs in set A and in set B are not consistent with being drawn from two different underlying distributions. Based on this result and to filter out stochastic fluctuations, we consider BBH mergers of set~A and set~B together in the following analysis.

\begin{figure}
  \center{
    \epsfig{figure=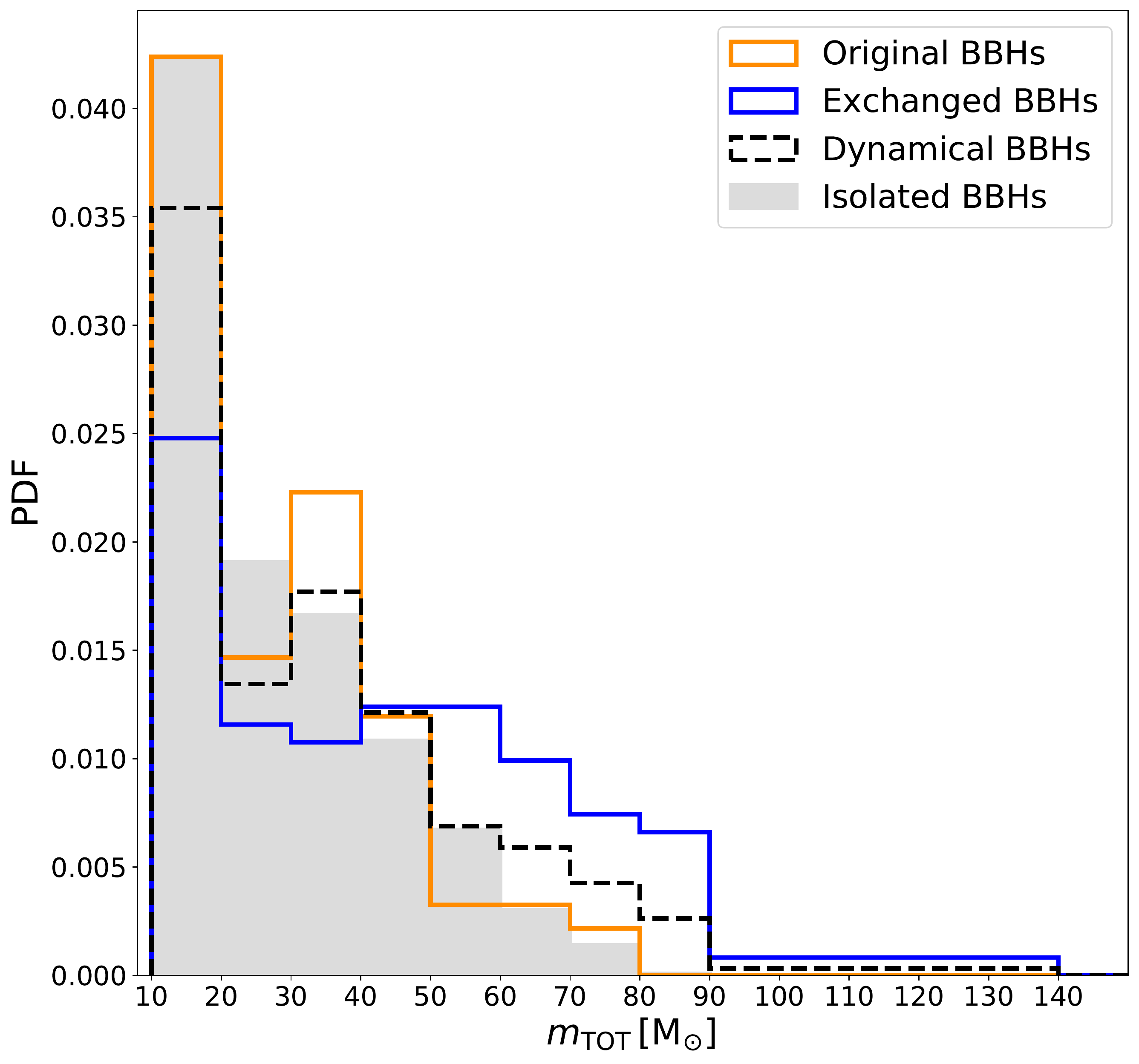,width=8.0cm}
    \caption{Distribution of total masses ($m_{\mathrm{TOT}}=m_1+m_2$) of merging BBHs. Set A and B are stacked together. Orange solid line: original BBHs; blue solid line: exchanged BBHs; black dashed line: all dynamical BBHs (original+exchanged); gray filled histogram: isolated BBHs.}\label{fig:mtot}} 
\end{figure}

\begin{figure}
  \center{
    \epsfig{figure=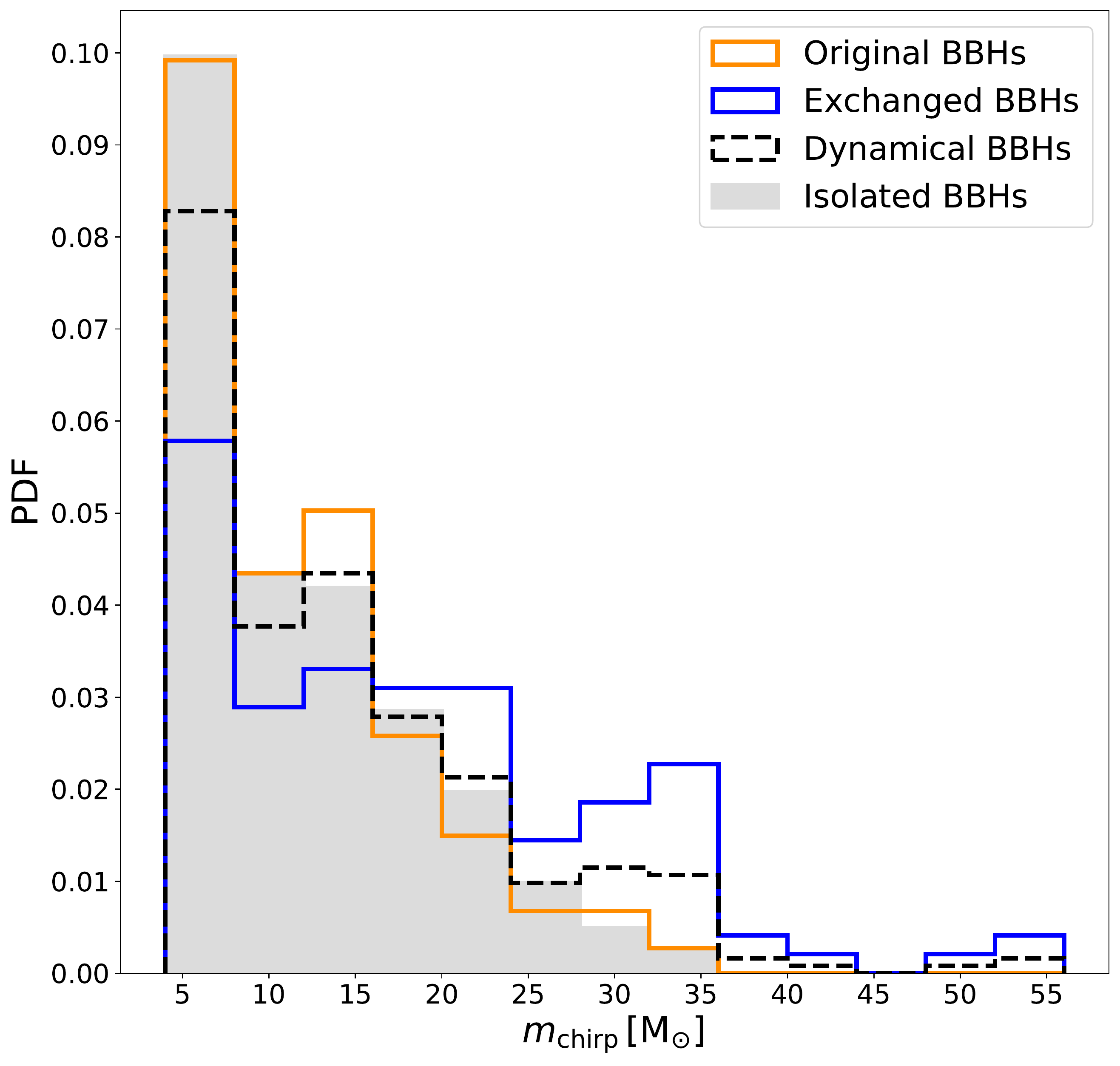,width=8.0cm}
    \caption{
    Same as Figure \ref{fig:mtot}, but for the distribution of chirp masses $m_{\mathrm{chirp}}=(m_1\,{}m_2)^{3/5}(m_1+m_2)^{-1/5}$ of merging BBHs.\label{fig:mchirp}}
}
\end{figure}
  
\begin{figure}
  \center{
    \epsfig{figure=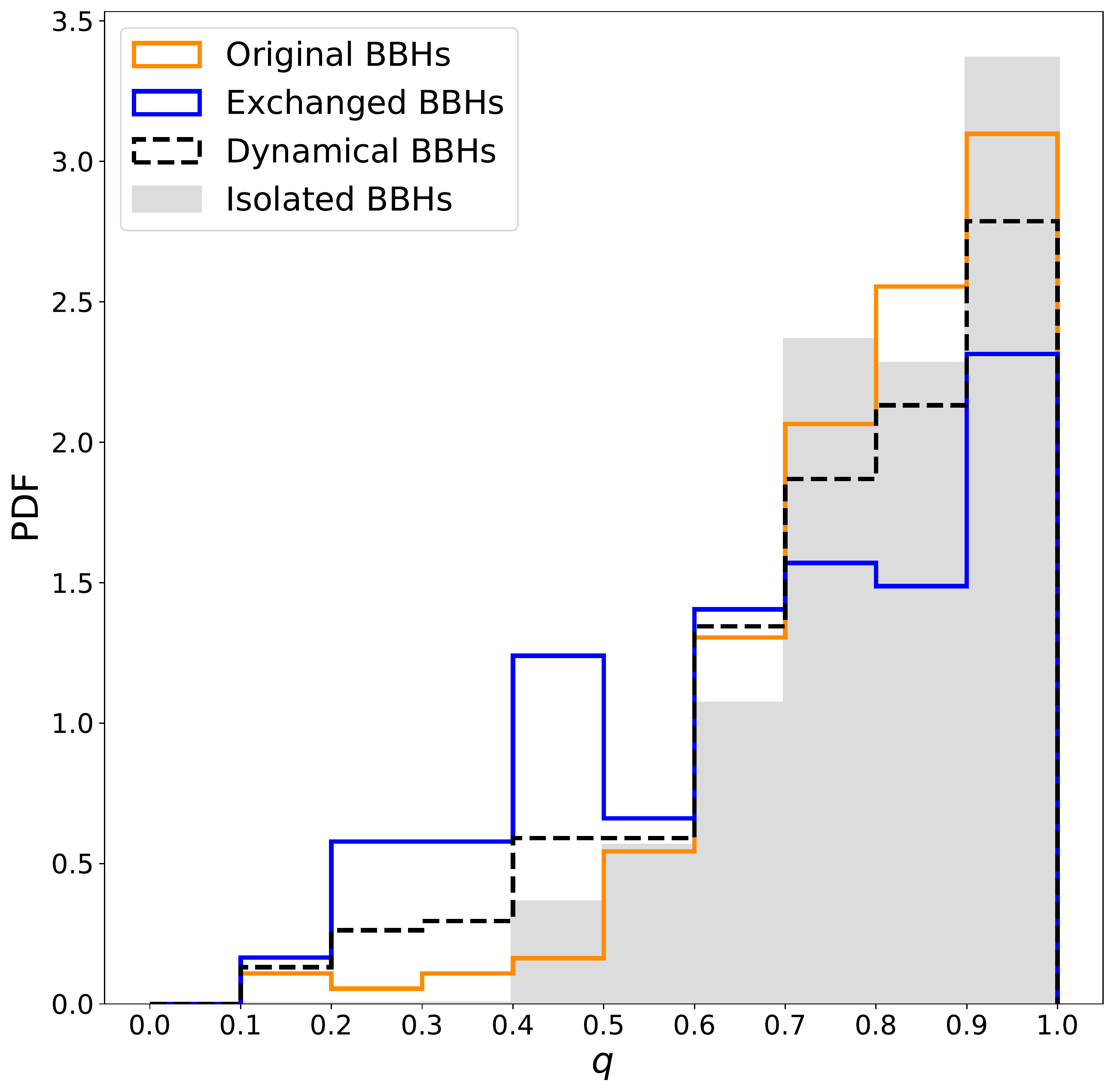,width=8.0cm}
    \caption{
    Same as Figure \ref{fig:mtot}, but for the distribution of mass ratios $q=m_2/m_1$ of merging BBHs.\label{fig:q}
    }
}
\end{figure}

\begin{figure}
  \center{
    \epsfig{figure=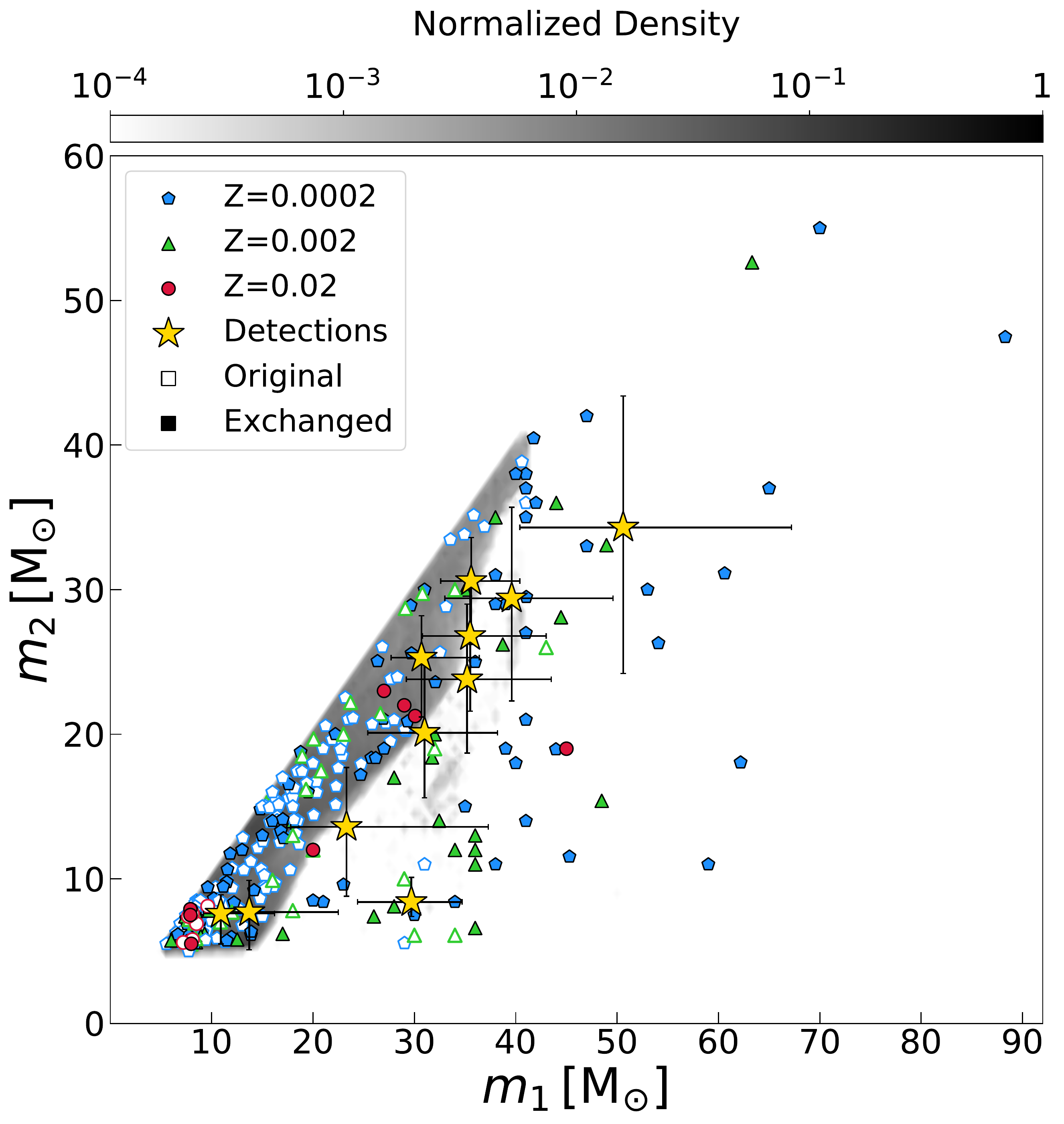,width=8.0cm}
    \caption{\label{fig:m1m2}
    Mass of the primary BH ($m_1$) versus mass of the secondary BH ($m_2$) of merging BBHs. Set A and B are stacked together. Empty symbols: original BBHs; filled symbols: exchanged BBHs. Blue, green and red symbols represent $Z=0.0002$, 0.002 and 0.02, respectively. Filled contours (with gray colour map): isolated BBHs. Yellow stars with error bars: LVC BBHs 
    [GW150914 \citep{abbottGW150914}, GW151012 \citep{abbottO1}, GW151226 \citep{abbottGW151226}, GW170104 \citep{abbottGW170104}, GW170608 \citep{abbottGW170608}, GW170729 \citep{abbottO2}, GW170809 \citep{abbottO2}, GW170814 \citep{abbottGW170814}, GW170818 \citep{abbottO2}, GW170823 \citep{abbottO2}, GW190412 \citep{abbottGW190412}]. 
    Error bars indicate 90\% credible levels.}
    }

\end{figure}

\begin{table}
\begin{center}
\caption{\label{tab:heavy} List of the BBH mergers with primary mass $m_1\ge{}45$ M$_\odot$ in our simulations.} \leavevmode
\begin{tabular}[!h]{cccccc}
\hline
$m_1$ [M$_\odot{}$] & $m_2$ [M$_\odot{}$] & $q$ & $Z$ & $t_{\mathrm{delay}}$ [Gyr] & Set 
\\
\hline
$88.3$ & $47.5$ & 0.54 & $0.0002$ & $0.046$ & A\\
$70.0$ & $55.0$ & 0.79 & $0.0002$ & $1.679$ & B\\
$65.0$ & $37.0$ & 0.57 & $0.0002$ & $0.0324$ & A\\
$63.3$ & $52.6$ & 0.83 & $0.002$ & $11.008$ & B\\
$62.2$ & $18.0$ & 0.29 & $0.0002$ & $0.264$ & B\\
$60.6$ & $31.1$ & 0.51 & $0.0002$ & $5.876$ & A\\
$59.0$ & $11.0$ & 0.19 & $0.0002$ & $0.499$ & A\\
$54.1$ & $26.3$ & 0.49 & $0.0002$ & $0.253$ & A\\
$53.0$ & $30.0$ & 0.57 & $0.0002$ & $7.0178$ & A\\
$49.0$ & $33.1$ & 0.68 & $0.002$  &  $0.505$ & B\\
$48.5$ & $15.4$ & 0.32 & $0.002$  & $0.117$ & A\\
$47.0$ & $42.0$ & 0.89 & $0.0002$ & $0.0447$ & A\\
$47.0$ & $33.0$ & 0.70 & $0.0002$ & $0.437$ & B\\
$45.3$ & $11.5$ & 0.25  & $0.0002$ & $3.586$ & A\\
$45.0$ & $19.0$ & 0.42  & $0.02$   & $0.308$ & A\\
\hline
\end{tabular}
\end{center}
\begin{flushleft}
\footnotesize{Column~1: Mass of the primary BH ($m_1$); column~2: mass of the secondary BH ($m_2$); column~3: mass ratio ($q$); column~4: progenitor's metallicity ($Z$); column~5: delay time ($t_{\mathrm{delay}}$); column~6: simulation set.}
\end{flushleft}
\end{table}

Figures~\ref{fig:mtot}, \ref{fig:mchirp} and \ref{fig:q} show the total mass ($m_{\rm TOT}=m_1+m_2$), the chirp mass [$m_{\mathrm{chirp}}=(m_1\,{}m_2)^{3/5}(m_1+m_2)^{-1/5}$] and the mass ratio ($q=m_2/m_1$, where $m_1\ge{}m_2$) of merging BBHs, respectively. In these figures, the three metallicity samples and the two simulation sets are stacked together.

The total masses of dynamical BBH mergers range from $\sim{}10$ to $\sim{}140$~M$_\odot$, while the chirp masses span from $\sim{}4.8$ to $\sim{}55.8$ M$_\odot$. Mass ratios of order of one are most common, but the distributions reach a minimum value of $q\sim{}0.18$.

Exchanged BBHs reach significantly larger total masses and chirp masses and smaller values of $q$ than both original BBHs and isolated BBHs. The typical masses of original BBHs are similar to those of isolated BBHs. This confirms the results of \cite{dicarlo2019}, who considered only one metallicity ($Z=0.002$).

Figure~\ref{fig:m1m2} shows  the mass of the secondary BH ($m_2$) versus the mass of the primary BH ($m_1$), distinguishing between different metallicities. The most massive objects ($m_1>45$ Msun) form only at low metallicity ($Z=0.0002$, 0.002) and are exclusively exchanged BBHs.

Table~\ref{tab:heavy} shows the masses, metallicities and delay times of BBH mergers with primary mass $m_1\ge{}45$~M$_\odot$. All of them are exchanged BBHs and (according to our population-synthesis model) cannot form by isolated binary evolution. We choose this threshold of $45$ M$_\odot$, because \cite{abbottO2popandrate} indicate that the mass distribution of the primary BH in O1 and O2 LVC events is well approximated by models with no more than 1~\% of BHs more massive than 45~M$_\odot$. In our simulations, we show that these BBH mergers are impossible to form via isolated binary evolution, but can arise from dynamical exchanges in YSCs. These massive BBH mergers are $4.3$\% and $7.0$\% of all the BBH mergers we find in set~A and set~B, respectively. Most of them have mass ratios  different from one. 

\begin{figure}
  \center{
    \epsfig{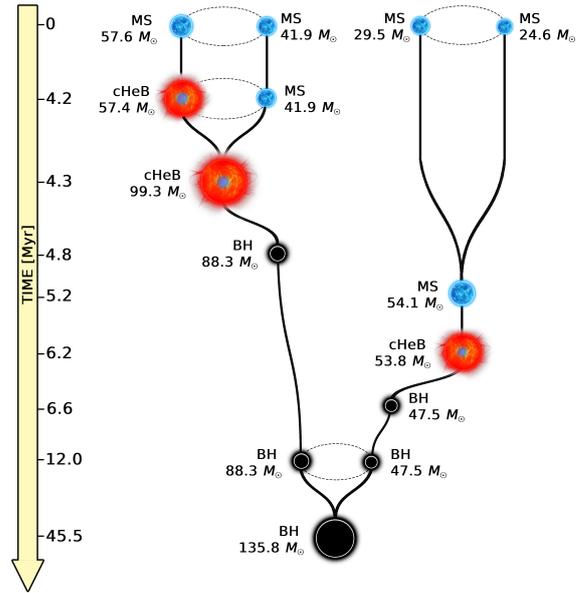}
    \caption{\label{fig:evolution}
     Evolution of the most massive BBH merger in our simulations. Blue stars  represent main sequence stars (with label MS); red stars with a blue core  represent core helium burning stars (label cHeB); black circles  represent black holes (label BH). The mass of each object is shown next to them. The time axis and the size of the objects are not to scale. The primary BH with $m_1=88.3$ \msun{} lies in the pair-instability mass gap. The merging BBH forms because of dynamical interactions. }
    }
\end{figure}

Figure \ref{fig:evolution} shows the evolution of the most massive BBH merger in our simulations, with a primary mass $m_1=88$ M$_\odot$ and a secondary mass $m_2=48$ M$_\odot$.  Both the primary and the secondary BH in this system form from the merger of two progenitor stars and become bound by exchange.  The mass of the primary BH is within the pair instability mass gap. This happens because the merger between a core helium burning (cHeB) star and a main sequence (MS) star produces a new cHeB star with a large hydrogen envelope and with a helium core below the threshold for (pulsational) pair instability (see \citealt{dicarlo2020} for further details). The merger between the 57.4 M$_\odot$ cHeB and the 41.9 M$_\odot$ MS is triggered by a dynamical encounter. If we simulate a binary with the same initial conditions using the stand-alone version of {\sc mobse} (i.e. without dynamical perturbations), the binary does not merge at 4.3 Myr and leaves a smaller remnant.

The yellow stars in Figure~\ref{fig:m1m2} show the 10 BBHs detected by the LVC during O1 and O2 \citep{abbottO2} plus GW190412, the first published BBH merger of O3 and the first event showing evidence of unequal mass components  \citep{abbottGW190412}. Our simulated BBH mergers match all O1--O2 BBHs including GW170729. GW170729, the most massive event detected in O1 and O2,  is consistent only with BBHs formed in YSCs (mostly exchanged BBHs): our models cannot form GW170729 via isolated binary evolution, even at the lowest considered metallicity. This result strongly favours a dynamical formation for GW170729. Even GW190412 can be matched only by dynamical BBHs born from metal-poor progenitors, because isolated binaries can hardly account for its mass ratio in our models.

\begin{figure}
  \center{
    \epsfig{figure=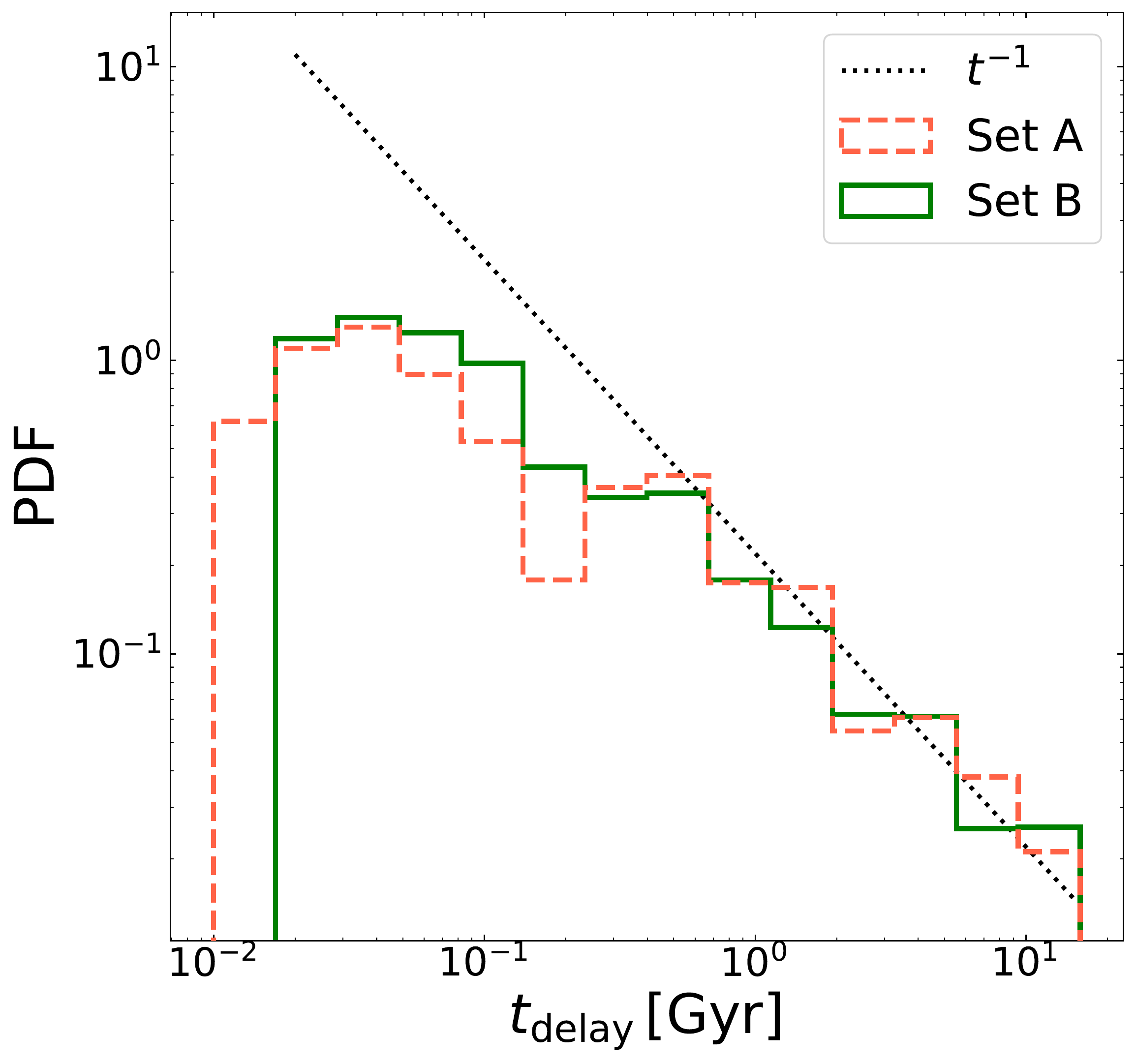,width=8.0cm}
    \caption{\label{fig:tdel}
    Distribution of delay times $t_{\mathrm{delay}}$ of merging BBHs. Orange dashed line: set~A; green solid line: set~B. Dotted black line: scaling as $dN/dt\propto{}t^{-1}$.}
    }
\end{figure}

Figure~\ref{fig:tdel} shows the distribution of delay times for our simulated BBHs. We find no significant differences between the delay time distribution of set~A and set~B (see Table~\ref{tab:stat}). The two distributions are broadly consistent with $dN/dt\propto{}t^{-1}$ \citep{dominik2012} if $t_{\rm delay}\gtrsim{}400$ Myr, but  bend  with respect to this scaling at shorter times. As a result, the overall distributions are not consistent with $\propto{}t^{-1}$, unless we neglect delay times shorter than 400 Myr. 

\begin{figure}
  \center{
    \epsfig{figure=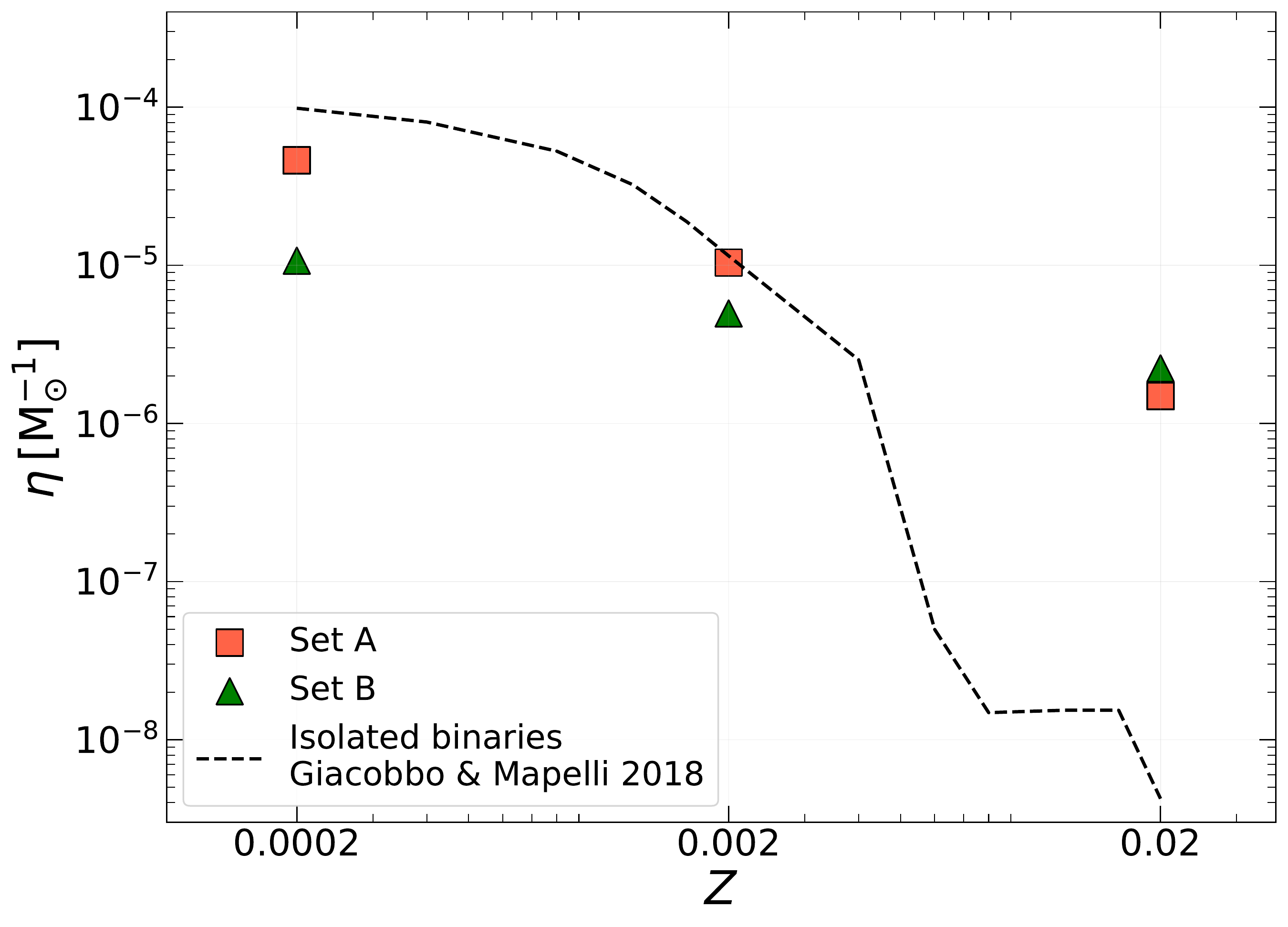,width=8.0cm}
    \caption{\label{fig:mergeff}
    Merger efficiency $\eta(Z)$, defined as the number of mergers per solar mass, as a function of metallicity. The black dashed line shows the values from \protect\cite{giacobbo2018b}. Orange squares and green triangles refer to set A and set B, respectively. 
}
}
\end{figure}

\subsection{Merger efficiency and local merger rate}
Figure~\ref{fig:mergeff} shows the merger efficiency $\eta{}(Z)$ defined as in \cite{giacobbo2018b}:
\begin{equation}
\eta{}(Z)=\frac{N_{\rm TOT}(Z)}{M_\ast(Z)},
\end{equation}
where $N_{\rm TOT}(Z)$ is the total number of BBHs (formed at a given metallicity) with delay time shorter than the Hubble time, while $M_\ast{}(Z)$ is the total initial stellar mass of the simulated population at a given metallicity. For isolated binaries  \citep{giacobbo2018b}, $M_\ast(Z) = M_{\ast,\mathrm{sim}}(Z)/(f_{\mathrm{bin}}\, f_{\mathrm{corr}})$, where $M_{\ast,\mathrm{sim}}(Z)$ is the total initial mass of the simulated binaries, $f_{\mathrm{bin}} = 0.4$ accounts for the fact that we simulated only binaries and not single stars, and $f_{\mathrm{corr}}$ accounts for the missing low-mass stars between 0.1 and 5 \msun{}. 
The merger efficiency is a useful quantity to understand the impact of stellar metallicity on the merger rate of binary compact objects. 

The most remarkable difference between isolated BBHs and dynamical BBHs is that, at solar metallicity ($Z=0.02$), the merger efficiency of the latter is higher by two orders of magnitude than the merger efficiency of the former. In YSCs, exchanges lead to the formation of BBHs and dynamical encounters harden existing massive binary stars, even at solar metallicity. In contrast, isolated BBH mergers are much rarer at solar metallicity, because stellar winds are efficient: the vast majority of massive stars become Wolf-Rayet stars before they can start a Roche lobe episode and do not undergo a common envelope phase; hence, most of the isolated BBHs which form at solar metallicity are too wide to merge within a Hubble time \citep{giacobbo2018b}.

From the merger efficiency $\eta(Z){}$, we can estimate the local merger rate density $\mathcal{R}_{\text{BBH}}$, as already described in \cite{santoliquido2020}:


\begin{eqnarray}
\label{eq:rate}
   \mathcal{R}_{\text{BBH}} = \frac{1}{t_{\rm lb}(z_{\text{loc}})}\int_{z_{\rm max}}^{z_{\text{loc}}}\psi(z')\,{}\frac{{\rm d}t_{\rm lb}}{{\rm d}z'}\,{}{\rm d}z' \times{}\nonumber{}
   \\
   \int_{Z_{\rm min}(z')}^{Z_{\rm max}(z')}\eta{}(Z)\,{}\mathcal{F}(z',z_{\text{loc}}, Z)\,{}{\rm d}Z,
\end{eqnarray}
where $t_{\rm lb}(z_{\text{loc}})$ is the look-back time evaluated in the local universe ($z_{\text{loc}}\leq 0.1$), $\psi(z')$ is the cosmic SFR density at redshift $z'$ (from \citealt{madau2017}), $Z_{\rm min}(z')$ and $Z_{\rm max}(z')$ are the minimum and maximum metallicity of stars formed at redshift $z'$ 
and $\mathcal{F}(z_{\text{loc}}, z', Z)$ is the fraction of BBHs that form at redshift $z'$ from stars with metallicity $Z$ and merge at redshift $z_{\text{loc}}$ normalized to all BBHs that form from stars with metallicity $Z$. To calculate the lookback time $t_{\rm lb}$ we take the cosmological parameters ($H_{0}$, $\Omega_{\rm M}$ and $\Omega_{\Lambda}$)  from \cite{ade2016}. We integrate equation~\ref{eq:rate} up to redshift $z_{\rm max}=15$, which we assume to be the epoch of formation of the first stars.

From equation~\ref{eq:rate} we obtain a local merger rate density $\mathcal{R}_{\rm BBH}\sim{}55$ and $\sim{}110$ Gpc$^{-3}$~yr$^{-1}$ for set A and B, respectively, by assuming that all the cosmic star formation rate occurs in YSCs like the ones we simulated in this paper. If we repeat the same procedure for the isolated BBHs, we find $\mathcal{R}_{\rm BBH}\sim{}50$ Gpc$^{-3}$~yr$^{-1}$. Set~B gives the highest local merger rate density, because it has a higher number of BBH mergers at solar metallicity (which is the dominant metallicity at low redshift) with relatively short delay times. Considering the small sample of BBH mergers at $Z=0.02$ (5 BBHs in set~A and 8 BBHs in set~B), the difference of a factor of 2 between the two local merger rates is likely due to stochastic fluctuations.

The inferred merger rates are upper limits, since we do not take into account infant mortality of YSCs \citep{brinkmann2017,shukirgaliyev2017}, we do not use an observation-based local number density of YSCs \citep{portegieszwart2000} and we assume that all stars form in YSCs like the ones we simulated in this paper. It is more likely that a fraction of
all mergers comes from YSCs and another fraction from isolated binaries, globular clusters or nuclear star clusters. In a follow-up paper (Bouffanais et al., in prep), we will try to constrain these
percentages based on current LVC results.

\section{Discussion}

\subsection{Merger efficiency: dynamical versus isolated BBHs}

Why the merger efficiency of dynamical BBHs is lower than that of isolated BBHs at low metallicity, but higher at high metallicity? This result springs from two opposite effects. On the one hand, dynamical encounters tend to break some BBHs, especially low-mass BBHs with a relatively large orbital separation (see e.g. \citealt{zevin2017} and \citealt{dicarlo2019}). On the other hand, dynamics enhances the merger of massive BBHs by exchanges and by hardening. The former effect tends to decrease the merger efficiency, while the latter tends to increase it. 

At solar metallicity ($Z=0.02$), the merger efficiency of isolated BBHs is drastically low (2--3 orders of magnitude lower than at $Z\leq{}0.002$). 
This implies that, at solar metallicity, even if dynamics ionizes all the low-mass original BBHs, this has no effect on the merger efficiency, because these low-mass original BBHs were not going to merge anyway. Thus, the loss of BBH mergers due to binary ionization/softening is minimum at high $Z$. In contrast, the few dynamical BBH mergers at high $Z$ all come from dynamical hardening and dynamical exchanges. The net effect is that the merger efficiency of dynamical BBHs is higher than that of isolated BBHs at solar metallicity. 

At low $Z$, the situation is inverted. Most of the mergers from isolated BBHs come from low-mass BBHs (see e.g. \citealt{giacobbo2018}). Hence, when dynamics suppresses the merger of these low-mass BBHs (by softening or ionization), it removes most of potential merging systems from the game. In metal-poor clusters, dynamical hardening and exchanges are efficient in forming massive BBHs and in  triggering their merger, but these massive binaries are not sufficiently numerous to compensate for the loss of low-mass mergers.
Hence, the net effect is that the merger efficiency of dynamical BBHs is lower than that of isolated BBHs at low $Z$.





There is also a difference between Set~A (dense clusters) and Set~B (loose clusters). At low $Z$, the merger efficiency of Set~A is a factor of $\sim{}5$ higher than that of Set~B, while at higher $Z$ the two sets have almost the same merger efficiency. The main reason for this difference is that, at low $Z$, where BH masses are higher, dynamical hardening and exchanges are more effective in the dense clusters of set A than in the loose clusters of set B. 


\subsection{When do the exchanges happen?}
Table~\ref{tab:types} shows that most  of the exchanged BBHs that merge within a Hubble time undergo their first exchange when the binary system is still composed of two stars, i.e. before the collapse of the primary component to a BH. The  percentage of  exchanges whose result is a binary composed of two stars is $\sim{}54$\% and $\sim{}72$\% for set~A and B, respectively. 

The percentage of exchanges that lead to the formation of a BH -- star binary is zero in set~B
and up to $\sim{}17$\% in set A. 
Finally, $\sim{}30$~\% of all exchanges that lead to BBH mergers happen when the two BHs have already formed.

\begin{figure}
  \center{
    \epsfig{figure=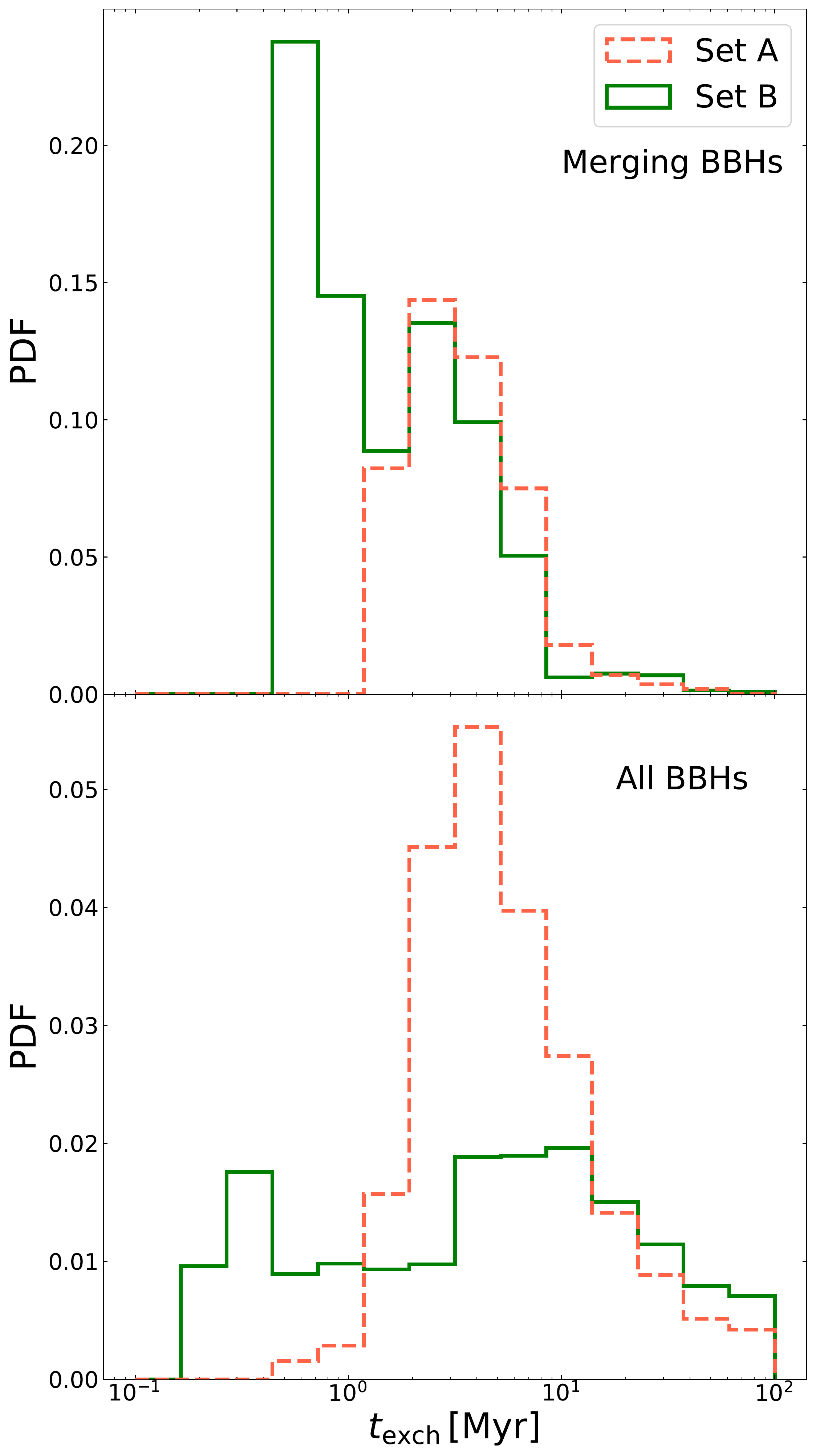,width=8.0cm}
    \caption{\label{fig:texch}
    Time when the first exchange took place for exchanged BBHs in set A (orange dashed line) and set B (green solid line). Top panel: merging BBHs only. Bottom panel: all BBHs.}}
\end{figure}

Figure~\ref{fig:texch} confirms this result: the dynamical exchanges that lead to the formation of merging systems happen in the first $\sim{}10$ Myr of the star cluster life. Most of these exchanges happen earlier ($t\ll{}1$ Myr) in the star clusters of set~B than in those of set~A ($t\sim{}2-3$ Myr).

This reflects a difference in the timescale for the collapse of the core of the cluster (hereafter, core collapse), because most interactions happen during core collapse. In set~B, the single sub-clumps of our fractal initial conditions undergo core collapse before they have completed the hierarchical assembly into the larger star cluster. Hence, most exchanges and dynamical interactions happen in this very early stage, $t<1$ Myr. In contrast, the  clusters of set~A are so dense that the sub-clumps hierarchically assemble to form one monolithic cluster before they undergo individual core collapse. Hence, the first core collapse in set~A is the collapse of the core of the global cluster at $t\sim{}2-3$ Myr. As already discussed  by \cite{2013MNRAS.430.1018F}, the build up and merger of massive binaries is suppressed if the sub-clumps collapse before the hierarchical assembly of the global cluster. Hence, we expect the binaries of set~A to start their dynamical activity later but to have more dynamical interactions with respect to the binaries of set~B.

\subsection{Integration time and merger rates}
We integrated all the simulated YSCs until their dissolution or for a maximum time  $t=100\,\mathrm{Myr}$. Would 
a longer integration time significantly affect the number of mergers? At the end of the simulations, our YSCs retain between 50\% and 70\% of their initial mass and $\sim60\%$ of the total BBHs. However, the vast majority of these in-cluster BBHs are  loose binaries ($\sim99.5\%$ of them have an orbital separation $a>10^2\,\mathrm{R_{\odot}})$ and would therefore require many strong dynamical interactions to harden and enter the GW regime. In a future work, we will integrate our clusters up to 1 Gyr to check the impact of the integration time on BBHs, but we do not expect it to significantly affect the number of mergers.

\subsection{Comparison with previous studies}

\cite{kumamoto2019} and \cite{kumamoto2020} evaluate the BBH merger rate from open clusters, whose masses and scales are comparable to our fractal YSCs. \cite{kumamoto2019} find that exchanges leading to BBH mergers happen mostly between stellar progenitors (before their collapse to BH), consistently with our results (see also \citealt{dicarlo2019}). Moreover, \cite{kumamoto2020} predict a local BBH merger rate density $\sim{}35$ Gpc$^{-3}$ yr$^{-1}$, similar to our result. \cite{banerjee2020} produced a set of simulations of more massive YSC, with masses between $10^4$ and $10^5$\msun and with lower binary fractions ($0<f_{\rm{bin}}<0.1$). \cite{banerjee2020} finds a mass spectrum of merging BBHs which is similar to our result; the main difference is that we find systems with lower mass ratios. Moreover, while 97\% of our BBH mergers take place outside the YSC, most of the mergers in \cite{banerjee2017} and \cite{banerjee2020} happen inside the cluster, likely because of the higher star cluster mass in these studies with respect to our simulations. 

These results for both YSCs \citep{dicarlo2019,dicarlo2020} and open clusters \citep{banerjee2010,ziosi2014,banerjee2017,banerjee2018} remark a crucial difference with respect to globular clusters (e.g. \citealt{portegieszwart2000,morscher2015,rodriguez2015,rodriguez2016,rodriguez2018,askar2017}) and nuclear star clusters (e.g. \citealt{antonini2016,arcasedda2019}). Globular  and nuclear clusters are significantly more long-lived than open  and young clusters. Hence, BBHs born in the former clusters have more time to harden by gravitational encounters and to undergo exchanges before they merge. This is expected to boost the merger efficiency per globular/nuclear cluster. On the other hand, most globular clusters formed $\sim{}12$ Gyr ago; hence, their contribution to the {\emph{local}} merger rate density is relatively small ($<20$ Gpc$^{-3}$ yr$^{-1}$, e.g. \citealt{askar2017,rodriguez2018b}). 

In contrast, YSCs are short-lived, but form all the time across cosmic history. Thus, they might have a larger cumulative effect on the local merger rate density of BBHs. Moreover, YSCs are the main birth-place of massive stars, and, when they are disrupted by gas evaporation or by the tidal field, they release their stellar content into the field. Thus, a large fraction of the field binaries might have formed in a YSC and might have taken part in dynamical encounters before their ejection/evaporation \citep{kruijssen2012}. 

A further difference between BBHs born in globular clusters and YSCs is the location of the mergers. About half of BBHs born in globular clusters are expected to merge inside the cluster \citep{banerjee2017,rodriguez2018,samsing2018,zevin2019}. In contrast, $\sim{}97$\% of our merging BBHs  reach coalescence after they were ejected from the YSC, because of the low escape velocity and of the short lifetime of these systems. Hence, most BBHs born in YSCs merge in the galactic field and might represent a large fraction of field mergers.


\begin{table}
\begin{center}
\caption{\label{tab:types} Progenitors of  exchanged BBH mergers at the time of the first exchange} \leavevmode
\begin{tabular}[!h]{cccc}
\hline
Set & $f_{\mathrm{star-star}}$ & $f_{\mathrm{star-BH}}$ & $f_{\mathrm{BBH}}$ 
\\
\hline
YSC & $57$\% & $12$\% & $31$\% \\
A & $54$\% & $14$\% & $32$\% \\
B & $72$\% & $0$\% & $28$\% \\
\hline
A02 & $0$\% & $0$\% & $100$\% \\
A002 & $45$\% & $9$\% & $46$\% \\
A0002 & $60$\% & $17$\% & $23$\% \\
B02 & $67$\% & $0$\% & $33$\% \\
B002 & $60$\% & $0$\% & $40$\% \\
B0002 & $80$\% & $0$\% & $20$\% \\
\hline
\end{tabular}
\end{center}
\begin{flushleft}
\footnotesize{Column~1: Simulation set; column~2: percentage of  exchanged BBH mergers in which the result of the first exchange is a star-star binary ($f_{\mathrm{star-star}}$); column~3: percentage of exchanged BBH mergers in which the result of the first exchange is a star--BH binary ($f_{\mathrm{star-BH}}$); column~4: percentage of exchanged BBH mergers in which the outcome of the first exchange is already a BBH ($f_{\mathrm{BBH}}$).}
\end{flushleft}
\end{table}

\section{Conclusions}
We have investigated the formation of BBH mergers in young star clusters (YSCs) with different metallicity, from $Z=0.0002$ to $Z=0.02$, by means of N-body simulations, coupled with the binary population-synthesis code {\sc{mobse}} \citep{giacobbo2018,dicarlo2019}. We probe two different density regimes for YSCs: dense clusters (set~A, i.e. clusters with half mass radius following the \citealt{marks2012} relation, corresponding to a density $\rho_{\rm h}\ge{}3.4\times10^4$ M$_\odot$ pc$^{-3}$) and loose clusters (set~B, i.e. clusters with half-mass radius $r_{\rm h}=1.5$ pc, corresponding to a density $\rho_{\rm h}\ge{}1.5\times10^2$ M$_\odot$ pc$^{-3}$, depending on star cluster mass).

We have shown that BHs and BBHs can reach higher masses at lower metallicity ($Z\leq{}0.002$) with respect to solar metallicity (Fig.~\ref{fig:mhist}). In our simulations, we can form IMBHs as massive as $\sim{}320$ M$_\odot$, through multiple stellar collisions. Stellar collisions also allow the formation of BHs with mass in the pair-instability mass gap \citep{dicarlo2020} even at solar metallicity, although their incidence is much higher at low metallicity ($Z\leq{}0.002$). We find that $\sim{}6$ \% ($\sim{}2$ \%) of all BHs formed at $Z=0.0002$ ($Z=0.002$) have mass $m_{\rm BH}>60$ M$_\odot$, while at solar metallicity ($Z=0.02$) the percentage is $<1$~\% in both set~A and B.

 The mass function of BHs and BBHs does not show significant differences between loose clusters (set B) and dense clusters (set A). In particular, IMBHs form nearly with the same frequency in both loose and dense clusters.

We focus on the sub-sample of BBHs that merge within a Hubble time. About 60\% of them come from original binaries (i.e. binary stars that are already there in the initial conditions), while the remaining $\sim{}40$\% form from dynamical exchanges. Exchanges in YSCs mostly involve stars before they collapse to BHs, because of the short core-collapse timescale of YSCs ($<3$ Myr).

Exchanged BBH mergers reach higher total masses (up to $\sim{}140$ M$_\odot$) than original and isolated BBH mergers (maximum total mass $\sim{}80$ M$_\odot$, Fig.~\ref{fig:mtot}). The reason is that non-conservative mass transfer tends to reduce the maximum mass of BBH mergers in isolated and original binaries. Moreover, exchanged BBHs tend to have lower mass ratios ($q=m_2/m_1$) than original and isolated BBHs (Fig.~\ref{fig:q}). 

In our models, the most massive event reported by the LVC in O1 and O2, GW170729 \citep{abbottO2,abbottO2popandrate}, can be explained only with dynamical BBHs: almost all of them are exchanged BBHs and come from metal-poor progenitors (Fig.~\ref{fig:m1m2}). Even GW190412, the first unequal-mass BBH merger, can be explained only by BBHs born in YSCs: isolated binaries can hardly explain such extreme mass ratios, according to the models presented here.

The most massive BBH merger in our simulations has $m_{\rm TOT}\sim{}136$ M$_\odot$, primary mass $m_1\sim{}88$ M$_\odot$ and secondary mass $m_2\sim{}48$ M$_\odot$ (Table~\ref{tab:heavy}). The primary mass is inside the pair-instability mass gap and the total mass of the merger product classifies it as in IMBH. This system is more massive than all the O1 and O2 LVC BBHs.

The merger efficiency (i.e. the number of mergers divided by the total simulated mass) is about two orders of magnitude higher for dynamical BBHs than for isolated BBHs at solar metallicity (Fig.~\ref{fig:mergeff}). The main reason is that dynamical encounters and hardening trigger the merger of BBHs even at high metallicity, where binary evolution is unlikely to produce mergers. 

The main difference between loose  and dense clusters is the merger efficiency. At low metallicity, the merger efficiency of loose clusters is a factor of $\sim{}5$ lower than that of dense cluster, while at higher metallicity the merger efficiencies are comparable. Assuming that all the cosmic star formation rate takes place in YSCs, we find a local merger rate $\sim{}55$ ($\sim{}110$) Gpc$^{-3}$ yr$^{-1}$ in set~A (set~B), respectively. This shows that most BBH mergers might have originated in YSCs. Future studies will quantify the impact of YSCs on the total merger rate of BBHs, BHNSs and BNSs, based on the comparison with LVC observations.


\section*{Acknowledgments}
We are grateful to the anonymous referee for their careful reading of our manuscript and their useful comments. We thank Astrid Lamberts, Long Wang and Serena Banfi for useful comments. UNDC acknowledges financial support from Universit\`a degli Studi dell'Insubria through a Cycle 33rd PhD grant.
MM, NG, YB, SR, FS and AB 
acknowledge financial support from the European Research Council (ERC) under European Union's Horizon 2020 research and innovation programme, Grant agreement no. 770017 (DEMOBLACK ERC Consolidator Grant).
MS acknowledges funding from the European Union's Horizon 2020 research and innovation programme under the Marie-Sklodowska-Curie grant agreement No. 794393. AAT acknowledges support from JSPS KAKENHI Grant Numbers 17F17764 and 17H06360. UNDC and AAT also thank the Center for Interdisciplinary Exploration and Research in Astrophysics at Northwester University for its hospitality.
This work benefited from support by the International Space Science Institute (ISSI), Bern, Switzerland,  through its International Team programme ref. no. 393  {\it The Evolution of Rich Stellar Populations \& BH Binaries} (2017-18).

\section*{Data Availability}
The data underlying this article will be shared on reasonable request to the corresponding authors.

\bibliography{./bibliography}
\end{document}